\documentclass[aps,prl,twocolumn,a4paper,10pt,notitlepage,footinbib,superscriptaddress,showpacs]{revtex4-1}%
\usepackage[english]{babel}
\usepackage[T1]{fontenc}
\usepackage{endnotes}
\usepackage{amssymb,amsmath,amsfonts}
\usepackage{textcomp}
\usepackage{graphicx,color}
\usepackage[dvipsnames]{xcolor}
\usepackage[utf8x]{inputenc}
\usepackage{float}

\usepackage{tabularx}

\renewcommand{\vec}[1]{\boldsymbol{\mathrm{#1}}}%

\begin{document}

\renewcommand{\vec}[1]{\boldsymbol{#1}}
\newcommand{\mean}[1]{{\left< #1 \right>}}

\title{Topological Sieving of Rings According to their Rigidity}

\author{Stefano Iubini}
\affiliation{
Department of Physics and Astronomy, University of Padova, 
Via Marzolo 8, I-35131 Padova, Italy
}

\author{Enzo Orlandini}
\email{orlandini@pd.infn.it}
\affiliation{
Department of Physics and Astronomy and INFN, University of Padova, 
Via Marzolo 8, I-35131 Padova, Italy
}

\author{Davide Michieletto}
\affiliation{School of Physics and Astronomy, University of Edinburgh, Peter Guthrie Tait Road, Edinburgh, EH9 3FD, UK
}

\author{Marco Baiesi}
\affiliation{
Department of Physics and Astronomy and INFN, University of Padova, 
Via Marzolo 8, I-35131 Padova, Italy
}

\begin{abstract}
We present a novel mechanism for resolving the mechanical rigidity of
nanoscopic circular polymers that flow in a complex environment.  The
emergence of a regime of negative differential mobility induced by
topological interactions between the rings and the substrate is the
key mechanism for selective sieving of circular polymers with
distinct flexibilities.  A simple model accurately
describes the sieving process observed in molecular dynamics
simulations and yields experimentally verifiable analytical
predictions, which can be used as a reference guide for improving
filtration procedures of circular filaments. 
 The {\it topological sieving} mechanism we propose ought to be relevant also in
probing the microscopic details of complex substrates.
\end{abstract}

\maketitle

The sieving of fluctuating fibers or polymers according to some of their physical or topological properties is a key process in many research fields ranging from molecular biology~\cite{Andrews1964,Porath1968,Doenecke1975}, engineering~\cite{Chung2014} and polymer physics~\cite{Volkmuth1994,Alon1997,Han2000}. Beyond its theoretical appeal, the achievement of efficient separation techniques has far reaching industrial~\cite{Corma1997} and medical~\cite{Jee2005} applications and potentially broad impact on next-generation polymer-based materials~\cite{Ligon2017}.  
Most of the sieving techniques exploit the competition between external forcing, surface interactions or entropic trapping of the fibers due to obstacles~\cite{Viovy2000,Han2000}. This gives rise to unique transport properties that mostly depend on either the 
contour length~\cite{Calladine1997}, mass~\cite{Doenecke1975} or charge~\cite{Chung2014} of the filaments. Notably, traditional sieving techniques such as gel electrophoresis can even be employed to detect and separate biopolymers in different topological states~\cite{Stas,Weber2006a,Michieletto2015pnas}, e.g.~linear, circular, knotted or linked, and in turn provide the community with an irreplaceable tool to gain insight into a wide range of problems, from the packaging of DNA bacteriophages~\cite{Trigueros2001,Marenduzzo2009} to the topological action of certain classes of proteins in vivo~\cite{Bates2005,Baxter2011,Cebrian2014}.
Automated separation of polymers can also be obtained using microfluidic devices and a recent numerical study has shown that, due to hydrodynamic effects, circular and linear polymers can be separated by a Poiseuille flow within properly decorated micro-channels~\cite{Weiss2017}.
 
Despite the abundance of biologically and medically important circular biomolecules that may differ by their degree of rigidity, e.g., single and double stranded DNA plasmids~\cite{Calladine1997}, looped RNA and protein secondary structures~\cite{Micheletti2015,Sulkowska2012} or intasome-bound viral DNA~\cite{Michieletto2018hiv}, there is a notable lack of studies aimed at investigating the effect that polymer rigidity may have on the transport properties of circular filaments within either structured fluids such as gels~\cite{Pernodet1997} or arrays of obstacles~\cite{Volkmuth1992,Rahong2014}.

Here we combine nonequilibrium analytical theories and large-scale Brownian dynamics simulations to address this problem by focusing on a model of semiflexible rings forced to move within a mesh of barriers with dangling ends (DE), i.e., fibers that are not part of a closed loop of the gel network~\cite{vieira2017,lee2018}. 

 We consider semiflexible rings made of beads of diameter $\sigma$ connected by FENE springs~\cite{Klenin2000} and interacting via a standard truncated and shifted Lennard-Jones potential. 
Besides fully rigid circular rings, we explore two values of the chain persistence length $l_p=20\sigma$ and $5\sigma$, see Supporting Information S1 and Figure S1. The model can then be related to various types of circular biomolecules with different rigidity.
Here we focus on rings with contour length $L= 100\sigma$. To fix the ideas, for $\sigma=2.5$nm, i.e. the typical double-stranded DNA thickness~\cite{Rybenkov1993}, this corresponds to plasmids of about $L=250{\rm nm} \simeq 700$ base pairs.

The structured medium through which the rings migrate is modeled as 
a sequence of layers with constant gap $l_{\rm gel}=80\sigma$ larger than radius of gyration of the rings and orthogonal to the direction $x$ of the force (see Figure~S1(b)). Each layer is a static square grid of beads decorated by DE of size $\ell< l_{\rm gel}$ oriented opposite to the external field. The DE mimic open strands that can be either naturally present in organic gels~\cite{Mickel1977,Levene1987,Cole2002} or artificially imprinted in microfluidic arrays~\cite{Rahong2014}. By assuming $\sigma = 2.5$nm, the size of the pores corresponds to $l_{\rm gel}=200$nm, comparable to those measured in a 5\% agarose gel~\cite{Pernodet1997}. Finally, the size of the beads forming the gel is set to $\sigma_g=10\sigma\simeq 25$nm, close to the average width of both agarose bundles (about 30nm)~\cite{Pernodet1997} and nanowires in artificial arrays of obstacles (about 20nm)~\cite{Rahong2014}.

To investigate the effect of the substrate geometry, we consider two spatial organizations of the layers: aligned and staggered by $l_{\rm gel}/2$ in the directions  orthogonal to the driving force (see Figure~S1(c)). Since the staggered organization displays trapping of rings at lower forces, we choose this arrangement as our default substrate, unless otherwise stated. Each monomer composing the rings is subject to a constant force $f$ in the $x$ direction and its motion through the medium is simulated by evolving the corresponding Langevin equation at fixed volume and constant temperature $T=1$ (NVT ensemble with Boltzmann constant $k_B=1$) using a molecular dynamics engine (LAMMPS)~\cite{Plimpton1995}. The results are reported as a function of the adimensional force $F = f \sigma / T$, while time is expressed in units of the characteristic time $\tau=36$ns (see Supporting Information S2 for details).

\begin{figure}[tb]
\includegraphics[width=8cm]{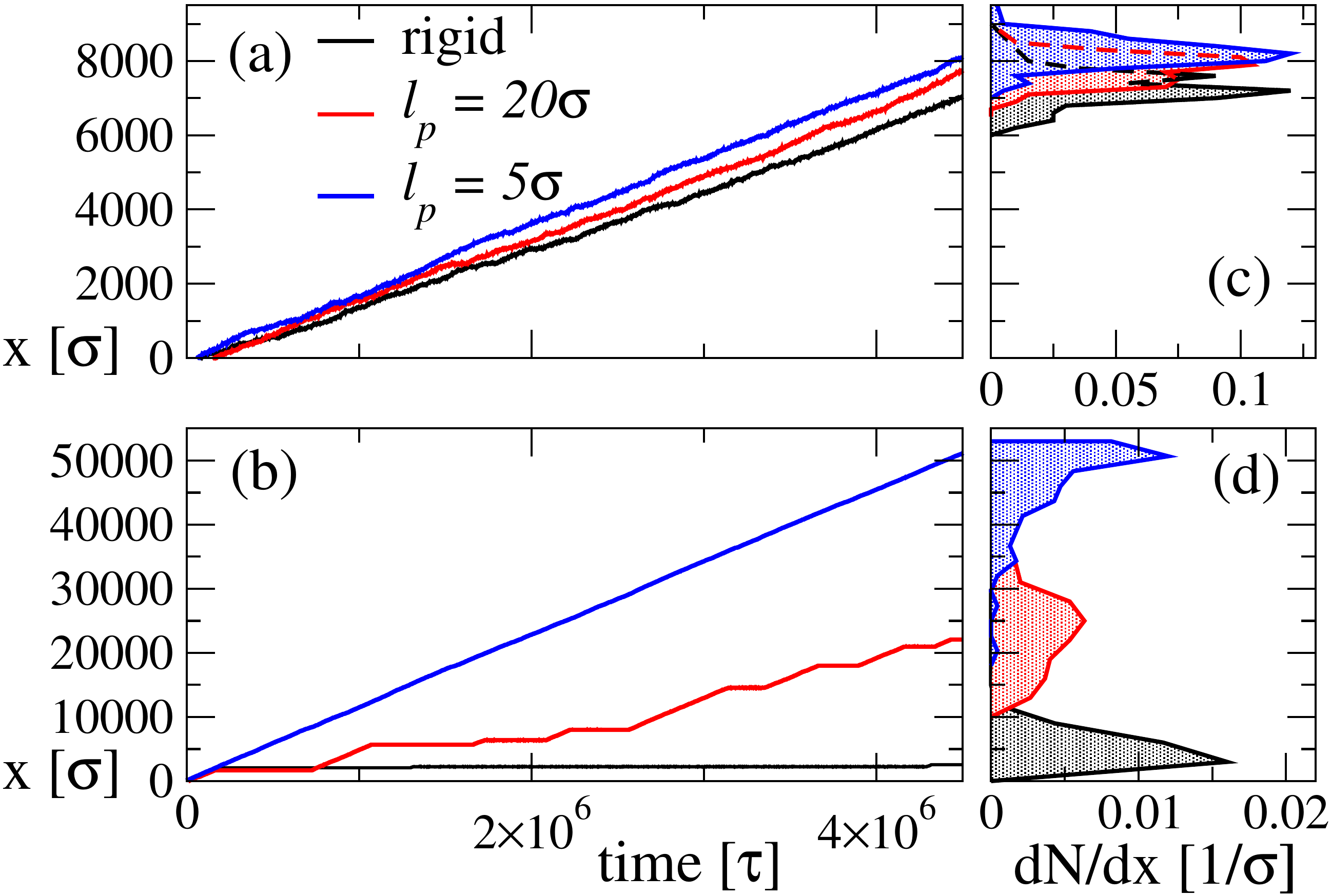} 
\caption{(a,b) Time dependence of the center of mass position of a ring along the 
force direction for $F=0.002$ (a) $F=0.012$ (b). Curves refer to a fully rigid ring (black) and
to a ring with persistence length $l_p = 20\sigma$ (red) and $l_p =5\sigma$ (blue).  
The DE length is $\ell=16\sigma$. Panel (c) and (d) are, respectively for case (a) and (b), the resulting late time spatial distributions of a sample of $100$ rings.
}
\label{fig:sieve}
\end{figure}

We first discuss the effect of chain rigidity on the transport properties of the rings. 
In Figure~\ref{fig:sieve}(a,b) we report typical trajectories of the center of mass along the field direction, respectively for $F=0.002$ and $F=0.012$, and for $\ell=2 \sigma_g = 16\sigma \simeq 40$nm. 
In each panel the three curves correspond to different rigidities. In Figure~\ref{fig:sieve}(a) the force is very low and no trappings are visible;  conversely, in Figure~\ref{fig:sieve}(b) trajectories are markedly separated. While most of flexible rings still migrate virtually undisturbed through the medium, those with $l_p=20\sigma$ display a more complex behavior, alternating runs (velocity $v>0$) with trappings ($v\simeq 0$). By visual inspection of Brownian dynamics trajectories, we associate trapping events with impalements in which rings are threaded by DE (see also  the movie in Supporting Information)~\cite{Michieletto2014softmatter}. These topologically trapped rings can only re-establish their motion by means of thermal fluctuations that transiently push them against the external field. The strong dependence of trapping and running typical times on rings rigidity gives rise to spatially separated classes of molecules which could be readily detected in electrophoretic experiments (see Figure~\ref{fig:sieve}(c,d)). This topological trapping cannot be seen for open filaments~\cite{Michieletto2014softmatter} but only for looped molecules.
In other words, while we expect topologically trivial polymers to display a rigidity-dependent mobility via classic Ogston sieving~\cite{Viovy2000}, these should not display trapping-driven topological sieving. Below we argue that this separation pathway may be important for polymers whose size is smaller than the gel pores, as they cannot be clearly separated via classic sieving~\cite{Viovy2000}.

\begin{figure}[tb]
\includegraphics[width=8cm]{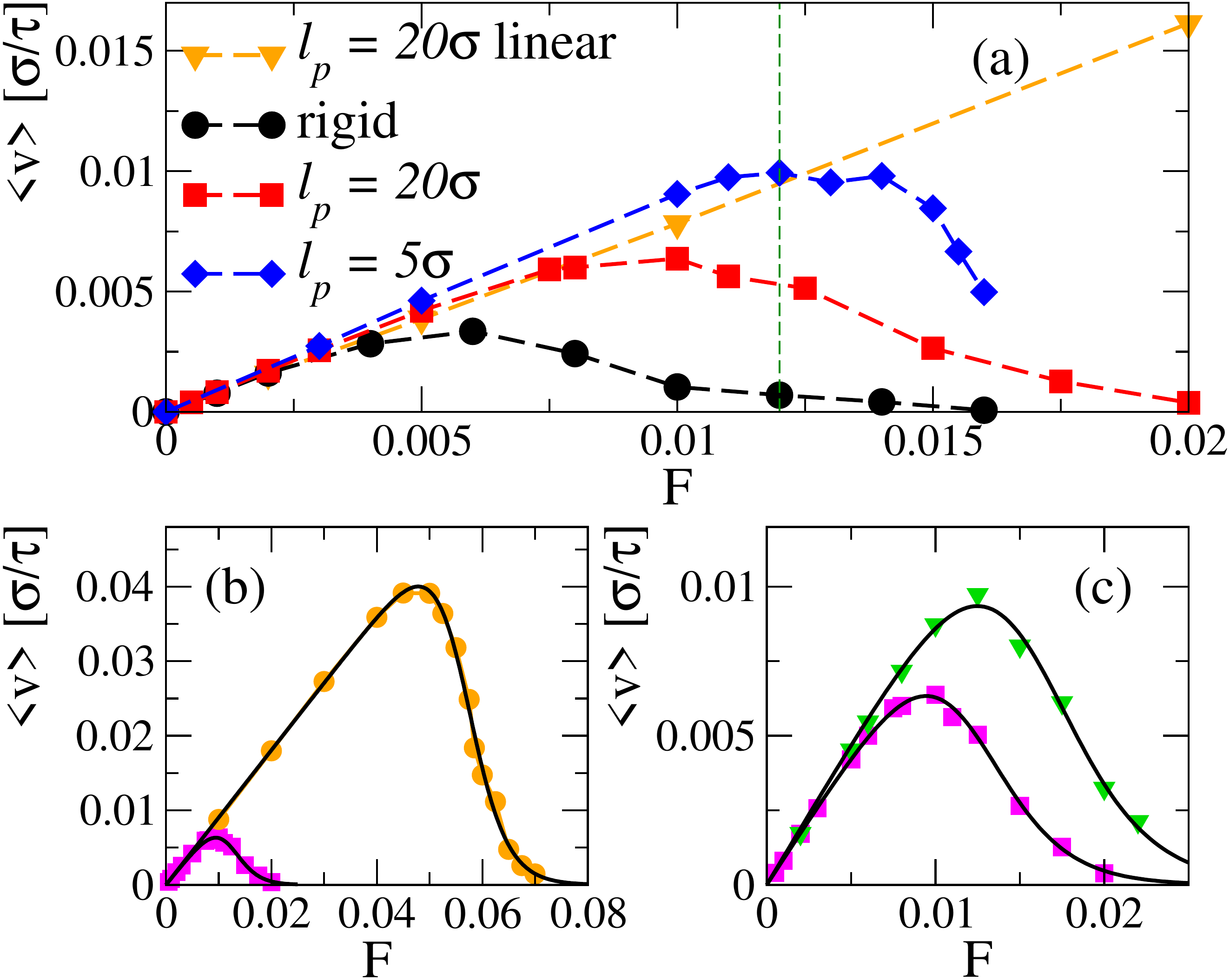} 
\caption{Examples of average ring velocity as a function of the driving $F$:
  (a) Data for rings with three rigidities (see legend), with DE length $\ell=16\sigma$. The vertical line indicates a region with good sieving properties. Dashed lines are guides to the eye.
  Data for linear polymers with $l_p=20\sigma$ and contour length $L= 99\sigma$ (orange triangles) show normal mobility, as short open polymers cannot be impaled by DE.
In (b) and (c) we compare data for $l_p=20\sigma$ and $\ell=16\sigma$ (magenta squares) with runs with (b) shorter DE length $\ell=8\sigma$ (orange circles) and (c) same parameters but in the cubic version of the gel (green triangles) rather than the staggered layers (default) version. In (b) and (c), continuous curves are fits according to (\ref{eq:vel}).
}
\label{fig:v}
\end{figure}

The polymer flow can be quantified by tracking the rings and by measuring their average speed.
As shown in Figure~(\ref{fig:v}), all systems display a non-monotonic response with a differential mobility $\mu_{\textrm{diff}} \equiv \frac{\partial \mean{v}}{\partial F} <0$
above a critical force $F_c$, i.e. a regime of  negative differential mobility  (NDM).
Additionally, the mobility of the rings markedly decreases with ring stiffness at large forces, yet this distinction is weaker or absent for small applied forces. We argue that in the latter regime, rings that become impaled by DE can easily escape by thermal fluctuations; 
conversely, for very large forces, the escape  probability of a trapped ring vanishes. 
Intriguingly, the most pronounced mobility difference is found at intermediate forces, thus suggesting that this regime may be the best candidate to achieve efficient and fast polymer separation. Finally, in Figure~\ref{fig:v}(a) we highlight that both the critical force $F_c$ and the response amplitude strongly depend on the ring flexibility; as we discuss below, this novel finding may be employed to refine current gel electrophoresis techniques.
As expected, no NDM emerges for linear polymers of the same length, see an example in Figure~\ref{fig:v}(a).

A minimal two-state model can account for the NDM and provide a simple description of the stationary state~\cite{Michieletto2014softmatter}. We assume that in the presence of a force $F$ the rings can either be trapped due to topological interactions (impalements) or running. In the running regime, the rings have an average non-zero velocity $v_R=\mu_R F$, where $\mu_R$ is the (positive) absolute mobility in the running state.
Simulations with no DE ($\ell=0$) show that $\mu_R$ is weakly dependent on $F$ and always lower than the value $\mu^{\rm free}_R=\sigma/\tau$ of a polymer in
solvent. Hereafter we use $\mu_R=0.9\,\sigma/\tau$ as a good approximation of the mobility of non-trapped rings.
Given $\mu_R$, the time to drift from one layer to the next is
$t_{\rm drift}=l_{\rm gel} / (\mu_R F)$, while the time scale to diffuse over a span $l_{\rm gel}$ is $t_D = l_g^2 / (2 \mu_R T)$. Diffusion thus is expected to dominate when $t_{\rm drift}>t_D$, i.e. (here using $T=1$) for $F<\tilde F=2/l_{\rm gel}=0.025$. Most of our simulations fall in this regime. Hence, the probability per unit time that a ring hits a DE and is impaled, namely the trapping rate $k_{\rm trap}$, is hereafter assumed to be independent on $F$. We checked that an additional parameter introducing a linear $F$-dependence of $k_{\rm trap}$ would not lead to visible improvements.

The transition rate from the trapped to the running state (or escape rate, $k_{\rm esc}$) takes into account an effective local energy barrier $\Delta E/ T = \alpha F$ which must be overcome by the ring when leaving the trapped state. By crudely approximating a ring as a point particle driven by $F$, the energy barrier would assume the simple form $\Delta E/T=F \ell/\sigma$ and the related escape rate $k_{\rm esc}$  would essentially depend on the product $F\ell$ only.
 However, due to the conformational entropy of the polymers, additional degrees of freedom, such as the ring persistence length $l_p$, can effectively enter into $\Delta E$ and produce more complicated responses.  To account for these effects, we write the escape rate in the more general form
\begin{equation}
  k_{\rm esc} = \psi \exp(-\alpha F)
  \label{eq:esc}
\end{equation}
where the parameters $\psi$ and $\alpha$ may depend on the ring rigidity and on $\ell$. 
By introducing the adimensional parameter $C=k_{\rm trap} / \psi$, we first derive the stationary probability of the running state
$p_R= \exp(-\alpha F)/ [ C + \exp(-\alpha F) ]$.  In turn,  the average ring velocity reads
\begin{equation}
\label{eq:vel}
\langle v \rangle = v_R  p_R
= \mu_R F \frac{\exp\left(-\alpha F \right)}{C+\exp\left(-\alpha F\right)} \, .
\end{equation}
The curve $\langle v(F)\rangle $ displays a region with NDM. More precisely, one finds $\mu_{\textrm{diff}} <0 $ for $F> F_c$, where $F_c$ solves the equation $C+e^{-\alpha F_c} -\alpha C F_c =0$.
Taking the limit of vanishing trapping, $C\rightarrow 0$ (or $\alpha\rightarrow 0$), this equation has no physical solution and the function $\langle v(F)\rangle $ becomes linear in $F$, with a constant positive $\mu_{\textrm{diff}}$. This clarifies that the origin of NDM has to be found in the topological interactions between rings and DE. 

In order to compare the analytical predictions with numerical simulations, we probe the nonequilibrium stationary states of the systems. In Figure~\ref{fig:v}(b) we compare the average velocity obtained from simulations with $l_p=20\sigma$ and for two DE lengths, and we fit the data with \eqref{eq:vel}. The fits give excellent results (curves in Figure~\ref{fig:v}(b)) also in the NDM regime. Note that the onset value $F_c$ of NDM drastically increases when $\ell$ is halved. This behavior is clearly related to the exponential dependence of $k_{\rm esc}$ on the energy barrier $\Delta E \approx \ell$.

Intriguingly, the trapping mechanism is also affected by spatial arrangement of DE in the substrate. As shown in Figure~\ref{fig:v}(c), systems differing only in the spatial position of the gel layers (aligned or staggered) display distinct responses to $F$. The staggered substrate reduces more quickly the velocity of the probes by increasing $F$, as the rings are more easily trapped by its exposed DE. In order words, we discover that $k_{\rm trap}$ can be tuned through the spatial organization of the gel, and it assumes larger values when layers are staggered, although the change in behavior is not as strong as the one consequent to a variation in $\ell$.

\begin{figure}[tb]
\includegraphics[width=8cm]{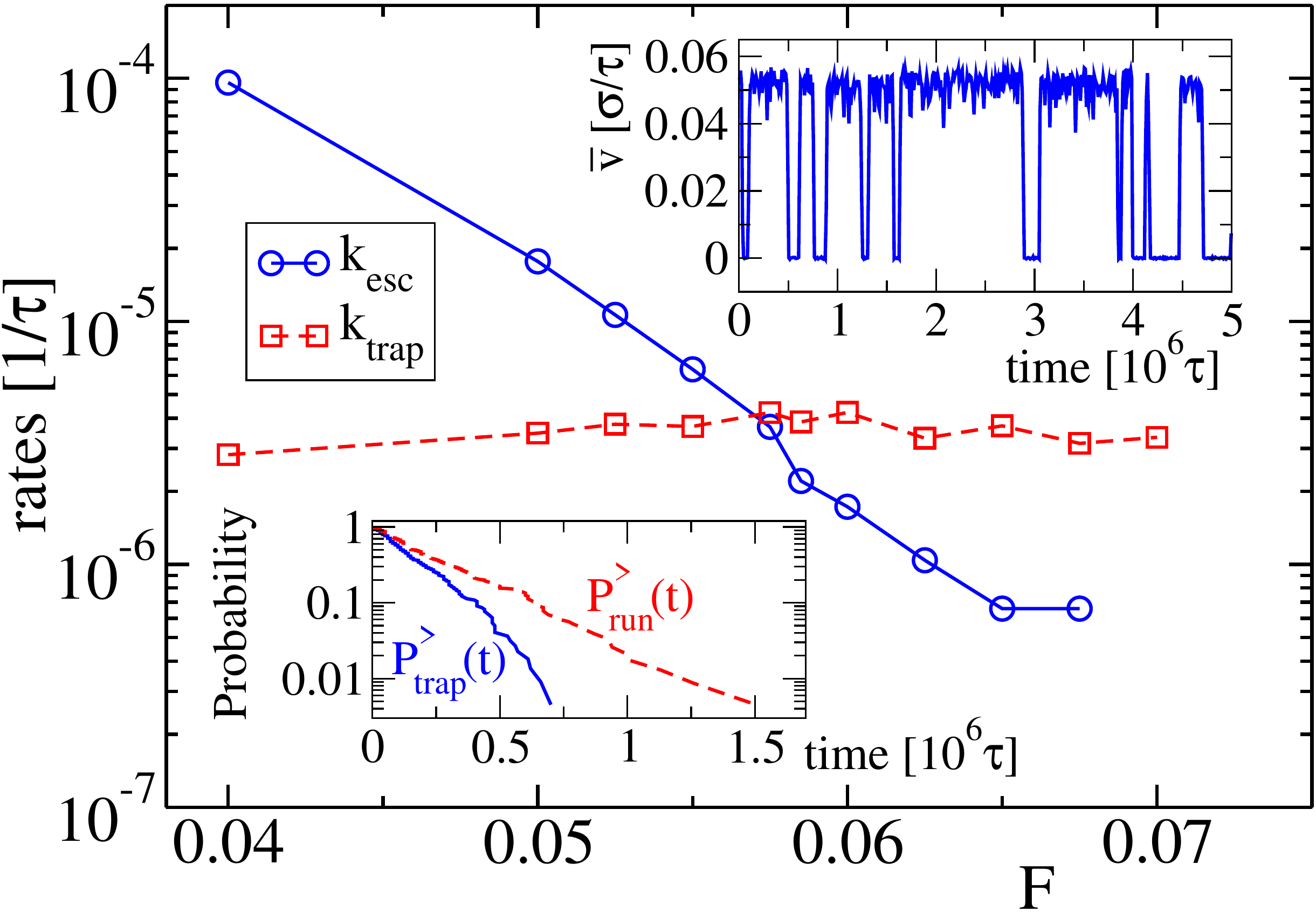} 
\caption{Rates $k_{\rm esc}$ (blue circles) and  $k_{\rm trap}$ (red squares) versus force $F$ for an ensemble of rings with $l_p=20\sigma$ in a gel with $\ell=8\sigma$. Lines are guides to the eye. Upper inset:  typical evolution of the  velocity $\bar{v}(t)$ of a ring for $F=0.055$. Lower inset: integrated probability distributions of trapped and running periods.
}
\label{fig:ex_time_ser}
\end{figure}

By tracking single-molecule trajectories, we can also explore the behavior of the trapping and escape rates for different NDM regimes and test whether they follow, respectively, $k_{\rm trap} \simeq$ constant and \eqref{eq:esc}. To this end, we first compute the
average velocity $\bar{v}(t)$ of a ring over small temporal windows of $10 \tau$, as shown in inset of Figure~\ref{fig:ex_time_ser}. 
This quantity displays reduced fluctuations with respect to the instantaneous ring velocity as well as a clear pattern of alternating running and trapped states, whose typical duration times can then be readily recorded.

\begin{figure}[ht]
\includegraphics[width=8.2cm]{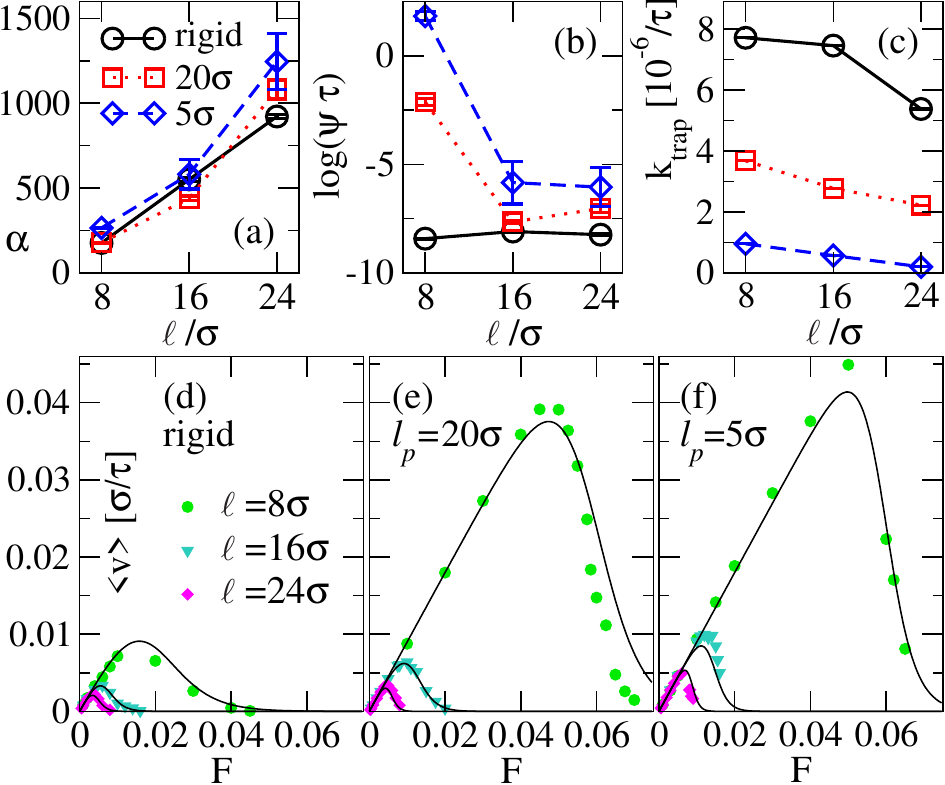}
\caption{The parameters  $\alpha$ (a), $\psi$ (b) and $k_{\rm trap}$ (c) for three plasmid rigidities, plotted vs  $\ell$. (d) For rigid rings, curves of average plasmid velocity as obtained from numerical simulations for different $\ell$'s (symbols, see caption) and from the two-states model with parameters shown in (a)-(c) (lines), and (e) the same for semiflexible polymers with $l_p=20\sigma$ and (f) $l_p=5\sigma$. 
}
\label{fig:param}
\end{figure}

Within the two-state model, both the probability distribution $P_{\rm trap}(t)$ of residence times in the trapped state and  $P_{\rm run}(t)$ of periods in the running state should follow an exponential decay with time. Hence, their integrated versions are
$P_{\rm trap}^>(t) \equiv \int_t^\infty  P_{\rm trap}(t') dt' = \exp(-k_{\rm esc}\, t)$
and
$P_{\rm run}^>(t)\equiv \int_t^\infty  P_{\rm run}(t') dt' = \exp(-k_{\rm trap}\, t)$,
and we can directly calculate the rates $k_{\rm esc}$ and $k_{\rm trap}$ by fitting their exponential scaling (see Figure~\ref{fig:ex_time_ser}).
By repeating the same analysis for systems with different rigidities and DE lengths, we are able to characterize precisely the parameters $\alpha$, $\psi$ and $k_{\rm trap}$, see Figure~\ref{fig:param}(a,b,c). 
In turn, we use the direct evaluation of these parameters from single-molecule trajectories to predict the mobility of the bulk for a given $\ell$, $l_p$ and $F$, with remarkably good results (Figure~\ref{fig:param}(d)-(f)). The sensitivity of the kinetic parameters to the ring rigidity observed for various $\ell$'s suggests the robustness of the sieving process also for more realistic environments
with heterogeneous dangling ends.

In summary, we have shown that a minimal model for ring polymers traveling through a complex environment with key features of a realistic gel, as the presence of DE, can capture the poorly explained empirical evidence of NDM for circular plasmids~\cite{Mickel1977,Akerman1998,Cole2002,Michieletto2014softmatter}.
We thus argue that electrostatic interactions, albeit present in real situations, may not be crucial here.
It is important to notice that in this study we have neglected hydrodynamic interactions.
  These can certainly affect the nonequilibrium dynamics and shape of confined polymers~\cite{Hsiao2016,Weiss2017} but they would not hinder the occurrence of NDM, which is mainly based on the topological interactions between the rings and the dangling ends. Preliminary simulations with hydrodynamic interactions confirm this expectation, see Supporting Information S4.

The onset of NDM may occur at distinct applied forces, depending on the flexibility of the rings, the typical size of dangling ends, and their position. 
Suitable protocols optimally exploiting  NDM can therefore be designed to efficiently separate circular biomolecules with different rigidity, such as RNA and DNA. 
Importantly,  topological sieving can separate molecules that are smaller than the size of the gel pores, a goal impossible to achieve with normal Ogston sieving.  

A fascinating consequence of these results is the possibility to use circular polymers drifting through a medium as probes for its microstructure. To this end, a predictive theoretical tool is necessary. Here we have shown that a simple two-state nonequilibrium theory captures remarkably well the sieving process.
The trapping and escaping rates can be directly measured by tracking single molecules drifting through a medium of unknown structure and may be used to directly characterize its complexity.

\paragraph{Acknowledgements -- }
We acknowledge support from Progetto di Ricerca Dipartimentale BIRD173122/17.
DM and EO acknowledge networking support by the COST Action CA17139. 

\providecommand*\mcitethebibliography{\thebibliography}
\csname @ifundefined\endcsname{endmcitethebibliography}
  {\let\endmcitethebibliography\endthebibliography}{}

\end{document}